\documentclass[aps,showpacs,superscriptaddress,twocolumn]{revtex4}
\usepackage{amsfonts}
\usepackage{amsmath}
\usepackage{amsthm}
\usepackage{amscd}
\usepackage{amssymb}
\usepackage{subfigure}
\usepackage{amsxtra}
\usepackage{bm}           
\usepackage{bbm}
\usepackage{graphicx}
\usepackage{epstopdf}

\def\bi#1\ei {\begin{itemize}#1\end{itemize}}
\def\bn#1\en {\begin{enumerate}#1\end{enumerate}}
\def\bea#1\eea {\begin{align}#1\end{align}}
\def\bean#1\eean {\begin{align*}#1\end{align*}}
\def\ben#1\een {\begin{equation*}#1\end{equation*}}
\def\be#1\ee {\begin{equation}#1\end{equation}}
\def\bes#1\ees {\begin{equation}\begin{split}#1\end{split}\end{equation}}
\def\bear#1\eear {\begin{eqnarray}#1\end{eqnarray}}
\def\bear#1\eear {\begin{eqnarray*}#1\end{eqnarray*}}

\newcommand{\beq}{\begin{equation}}
\newcommand{\eeq}{\end{equation}}

\newcommand{\ket}[1]{\ensuremath{|#1\rangle}}

\newcommand{\ketbra}[2]{\ensuremath{| #1 \rangle \langle #2 |}}
\newcommand{\mean}[1]{\ensuremath{\langle #1 \rangle}}

\newcommand{\Gd}{\Gamma^{\delta}}

\newcommand{\Den}{\mathcal{D}(\mathcal{H})}
\newcommand{\Hil}{\mathcal{H}}
\newcommand{\Lik}{\mathcal{L}_i(\sigma)}
\newcommand{\id}{\ensuremath{\mathbbm 1}}

\renewcommand{\qed}{\ensuremath{\hfill \blacksquare}}

\newcommand{\tr}[1]{\ensuremath{\mbox{Tr}\left( #1 \right)}}
\newcommand{\Tr}[1]{\ensuremath{\mbox{Tr}\left[ #1 \right]}}
\newcommand{\sx}{\ensuremath{\sigma_{x}}}
\newcommand{\sy}{\ensuremath{\sigma_{y}}}
\newcommand{\sz}{\ensuremath{\sigma_{z}}}

\newtheorem{thm}{Theorem}

\newtheorem{obs}[thm]{Observation}

\begin{document}

\title{Reliable Entanglement Verification}

\author{Juan Miguel Arrazola}
\author{Oleg Gittsovich}
\author{John Matthew Donohue}
\author{Jonathan Lavoie}
\author{Kevin J. Resch}
\author{Norbert L\"utkenhaus}

\affiliation{Institute for Quantum Computing and Department of Physics and Astronomy,
University of Waterloo, 200 University Avenue West,
N2L 3G1 Waterloo, Ontario, Canada}

\begin{abstract}
Any experiment attempting to verify the presence of entanglement in a physical system can only generate a finite amount of data. The statement that entanglement was present in the system can thus never be issued with certainty, requiring instead a statistical analysis of the data. Because entanglement plays a central role in the performance of quantum devices, it is crucial to make statistical claims in entanglement verification experiments that are reliable and have a clear interpretation. In this work, we apply recent results by M. Christandl and R. Renner \cite{ReliableTomo} to construct a reliable entanglement verification procedure based on the concept of confidence regions. The statements made do not require the specification of a prior distribution, the assumption of independent measurements nor the assumption of an independent and identically distributed (i.i.d.) source of states. Moreover, we develop numerical tools that are necessary to employ this approach in practice, rendering the procedure ready to be applied to current experiments.  We demonstrate this technique by analyzing the data of a photonic experiment generating two-photon states whose entanglement is verified with the use of an accessible nonlinear witness.
\end{abstract}


\date{\today}

\maketitle

\section{Introduction}

Entanglement plays an essential role in various quantum information processing tasks \cite{horodecki_review,curty04a,vidal03prl,guehne09a} and experimental verification of entanglement is crucial for testing and characterizing quantum devices such as sources and channels \cite{Nathan}. As these devices move closer to the realm of practical technologies, our ability to perform reliable entanglement verification tests becomes increasingly important. Correspondingly, many theoretical and experimental procedures for entanglement verification have been proposed (see \cite{horodecki_review,guehne09a} and references therein) and their improvement and development remains an active area of research \cite{gao2010experimental,moroder2012detection}.

Any entanglement verification procedure can be thought of as a series of measurements on a physical system followed by an analysis of the outcomes. The data obtained from these measurements is necessarily finite and therefore the claim that entanglement was present can never be issued with certainty. More precisely, there will always be a non-zero probability that the data was produced from a separable state, regardless of what the data may be. We are thus forced to provide statistical statements that quantify our confidence that entanglement was indeed present. Naturally, the procedure that leads to these statements should have a clear interpretation, should not rely on unwarranted assumptions about state preparation and be readily implementable in practice \cite{enk07a}.

The most widely used approach consists of computing the standard deviation of measured quantities and using these as error bars to specify the uncertainty of the reported values \cite{guehne09a}. However, there are several conceptual issues with this approach \cite{RobustErrorBars,RBK2010}, including the fact that it can lead to counter-intuitive results \cite{jungnitsch_increasing_2010} and is known to be inadequate to deal with nonlinear expressions \cite{EadieBook}. This strongly asks for better alternatives and consequently other approaches have been recently suggested (see e.g. \cite{BKEntanglement}).

In this paper, we apply the work of Christandl and Renner on quantum state tomography \cite{ReliableTomo} to formulate a reliable method for analyzing the data of entanglement verification experiments. As shown in Ref. \cite{ReliableTomo}, the method does not rely on the specification of a prior distribution of prepared states nor on the assumption that they are independent and identically distributed. Additionally, it is suitable for experiments performing arbitrary quantum measurements and the final statements have a clear and well-defined operational interpretation. The approach relies on the concept of \textit{confidence regions}: regions of state space that contain the true state with high probability \cite{ReliableTomo}.

Applying this method requires the specification of a region of state space for all possible measurement outcomes, an issue that is not dealt with directly in Ref. \cite{ReliableTomo}. In our work we provide a recipe to assign confidence regions to data obtained from entanglement verification experiments that rely on entanglement witnesses. This assignment requires the evaluation of a non-trivial inequality for which we specifically develop numerical techniques to efficiently calculate it, rendering the entire method ready to be applied to current experiments. We demonstrate this fact by experimentally producing a family of entangled two-photon states whose entanglement is verified by an accessible nonlinear witness (ANLW) \cite{NLW}.

The remainder of this paper is organized as follows. For the sake of completeness, we first briefly outline the framework introduced in Ref. \cite{ReliableTomo} and summarize some of its main results. We then proceed to illustrate the data analysis procedure that we build and elucidate the numerical tools that we develop to perform the necessary calculations. Finally, we describe the experimental setup and analyze the results with our technique.

\section{Confidence regions}\label{confreg}

We now provide an overview of the main results of Ref. \cite{ReliableTomo} and direct the reader to this work for further details.  We begin by considering a collection of $n+k$ quantum systems described by a state $\rho^{n+k}$, each system associated with a Hilbert space $\Hil$ of dimension $d$. The measurement is performed only on the first $n$ systems and is described by a general POVM consisting of a set $\{B_i\}$ of positive operators satisfying $\sum_i B_i=\id_{\Hil}^{\otimes n}$. In the case of independent measurements of each of the systems, each element $B_i$ will be a tensor product of $n$ positive operators acting on a single copy of the state. However, it must be clear that the formalism does not require this assumption: one should always think of this POVM as an arbitrary, generally collective measurement on $\Hil^{\otimes n}$. The role of the remaining $k$ systems is purely operational: the goal of the entanglement verification procedure is to make predictions about the state of these remaining systems. More precisely, we want to know if these systems belong to regions of state space that contain only entangled states. Note that $n$ is the \textit{number of runs} of the experiment, producing $n$ systems which are then measured and the outcomes analyzed to build the predictions.

Consider an experiment in which the predictions are made only for a subset of $k'$ subsystems, $k'<k$. It was noted in Ref. \cite{ReliableTomo} that in the limit of $k\rightarrow\infty$, the reduced state of the $n+k'$ subsystems $\rho^{n+k'}=\text{Tr}_{k-k'}({\rho^{n+k}})$ can always be described by a permutationally-invariant state of the form $\int P(\sigma)\sigma^{\otimes(n+k')}d\sigma$ \cite{DeFinetti}. This corresponds to the usual independent and identically distributed (i.i.d) case in which many copies of a true state $\sigma$ are prepared according to some initial probability distribution $P(\sigma)$. Thus, in the scenario of an experiment that can in principle be repeated an arbitrary number of times ($k\rightarrow\infty$) and predictions are made on a sample of $k'$ states, the above result in fact provides a justification of the i.i.d.\ assumption that is common in the literature. For convenience, we will adopt this point of view but remind the reader that the i.i.d.\ assumption is not necessary for the validity of the upcoming results \cite{ReliableTomo}.

The data analysis procedure we will employ is a mapping that assigns a particular region of state space to every possible measurement outcome. Crucially, this mapping must be specified before the experiment is carried out. The regions are deemed \textit{confidence regions} if they contain the true state with high, user-specified probability. More precisely, for all $i$, denote by $R(B_i)$ the region assigned to outcome $B_i$. This region will be a subset of the space of density matrices $\Den$  associated to $\Hil$. Then any prescribed region $R(B_i)$ is deemed a confidence region with confidence level $1-\epsilon$ if it satisfies the property
\begin{align}\label{confregions}
\text{Prob}_{B_i}\left[\sigma\in R(B_i)\right]\geq 1-\epsilon, \hspace{0.3cm}\forall \sigma,
\end{align}
where $\text{Prob}_{B_i}\left[\sigma\in R(B_i)\right]$ is the expected probability of success with respect to the distribution $\tr{\sigma^{\otimes n}B_i}$ of the measurement outcomes $B_i$. In this picture, statistical statements take the following form: ``We have applied a procedure that, with probability at least $1-\epsilon$, assigns a region containing the prepared state $\sigma$''. It is important to emphasize that this probability refers to the success of the procedure before any measurements are carried out: in the end, the original input state $\sigma$ is either definitely contained in the assigned region or not. The quantity $1-\epsilon$ should thus be interpreted as the confidence level of the statement that the state is contained in the assigned region. This statement is valid for all possible states and outcomes and does not depend on extra assumptions about state preparation nor on the prior distribution $P(\sigma)$. This fact makes the procedure reliable and robust even in the cryptographic scenario in which $\sigma$ might have been chosen maliciously \cite{ReliableTomo}. 

A main result of Ref. \cite{ReliableTomo} was to provide a criteria to determine whether a given mapping from outcomes to regions succeeds in constructing confidence regions. This result is summarized as follows. Firstly, for each measurement outcome define the function
\beq\label{mu}
\mu_i(\sigma)=\frac{1}{\mathcal{N}}\tr{\sigma^{\otimes n} B_{i}}=\frac{1}{\mathcal{N}}\Lik,
\eeq
where 
\beq
\mathcal{N}=\int_{\Den}\Lik d\sigma\nonumber
\eeq 
is a normalization constant. The function $\tr{\sigma^{\otimes n} B_{i}}$ is usually referred to as the \textit{likelihood function}, so that $\mu_i(\sigma)$ is simply its normalized version. Furthermore, let $\{\Gamma_i\}$ be a collection of subsets of $\Den$, where the number of these regions is equal to the number of POVM elements $\{B_i\}$. For each set $\Gamma_i$ define the enlarged set
\beq\label{GammaDelta}
\Gd_{i}=\{\sigma: \exists \sigma'\in\Gamma_i \text{ such that } F(\sigma,\sigma')\geq \sqrt{1-\delta^2}\},
\eeq
where $F(\sigma,\sigma')=\tr{\sqrt{\sqrt{\sigma}\sigma'\sqrt{\sigma}}}$ is the fidelity and 
\beq\label{delta}
\delta^2=\frac{2}{n}\left[\ln\frac{2}{\epsilon}+(d^2-1)\ln n\right].
\eeq
If for all possible outcomes $B_i$ it holds that
\beq
\int_{\Gamma_i}\mu_i(\sigma)d\sigma\geq 1-\frac{\epsilon}{c_{n,d}}\label{criteria}\\
\eeq
with
\beq
c_{n,d}=2 n^{(d^2-1)/2},
\eeq 
then the assigned regions $\Gd_{i}$ are confidence regions with confidence level $1-\epsilon$ (Corollary 1, \cite{ReliableTomo}). In equation \eqref{criteria}, $d\sigma$ is the Hilbert-Schmidt measure: the flat measure on the set of density matrices of dimension $d$ induced from the Haar measure on the set of pure states of dimension $d\times d$ \cite{InducedMeasures}. It must be noted that the polynomial factor $2 n^{(d^2-1)/2}$ \cite{ChristandlPrivate} is an improvement on the term appearing in Ref. \cite{ReliableTomo}.

The above condition \eqref{criteria} can be more conveniently cast by referring directly to the quantity $1-\int_{\Gamma_i}\mu_i(\sigma)d\sigma$ and making a direct comparison with the term $\epsilon/c_{n,d}$. This can be achieved by instead integrating over the complement regions $\overline{\Gamma_i}=\{\sigma: \sigma\notin \Gamma_i\}$. Therefore we define
\begin{align}
\epsilon_2(B_i,\Gamma_i)&:=\int_{\overline{\Gamma_i}}\mu_i(\sigma)d\sigma\nonumber\\
&=\frac{\int_{\overline{\Gamma_i}}\mathcal{L}_i(\sigma)d\sigma}{\int_{\Den}\mathcal{L}_i(\sigma)d\sigma}.
\end{align}
For convenience, we will drop the explicit dependence on $B_i$ and $\Gamma_i$ from $\epsilon_2(B_i,\Gamma_i)$ whenever it is not necessary, while keeping in mind that its value will depend on the measurement outcome and the region assigned to it. Condition \eqref{criteria} can then be more conveniently cast as
\beq\label{criteria2}
\epsilon_2\cdot c_{n,d}\leq \epsilon.
\eeq
In summary, the assigned regions $\{\Gamma_i\}$ determine whether criteria \eqref{criteria2} is satisfied for some fixed value of $\epsilon$ and whenever it is, the enlarged regions $\Gd_i$ constitute confidence regions. It is these latter regions that we assign to each individual outcome in our data analysis procedure.

It is very important to note the role played by the polynomial factor $c_{n,d}$ and the enlarging parameter $\delta$. Because the dimension of the Hilbert space $d$ is fixed for a given experiment and typically large, the factor $c_{n,d}$ will be a high-order polynomial in the number of runs $n$. Satisfying condition \eqref{criteria2} will require $\epsilon_2$ to be much smaller than the value of $\epsilon$ that quantifies the confidence of the procedure. This can be problematic for small $n$ but will play only a minor role for larger values because $\epsilon_2$ decreases exponentially in $n$ whenever the maximum of the function $\mu_i(\sigma)$ is contained in the region $\Gamma_i$ \cite{ReliableTomo}.

\begin{figure}
\includegraphics[width=\columnwidth]{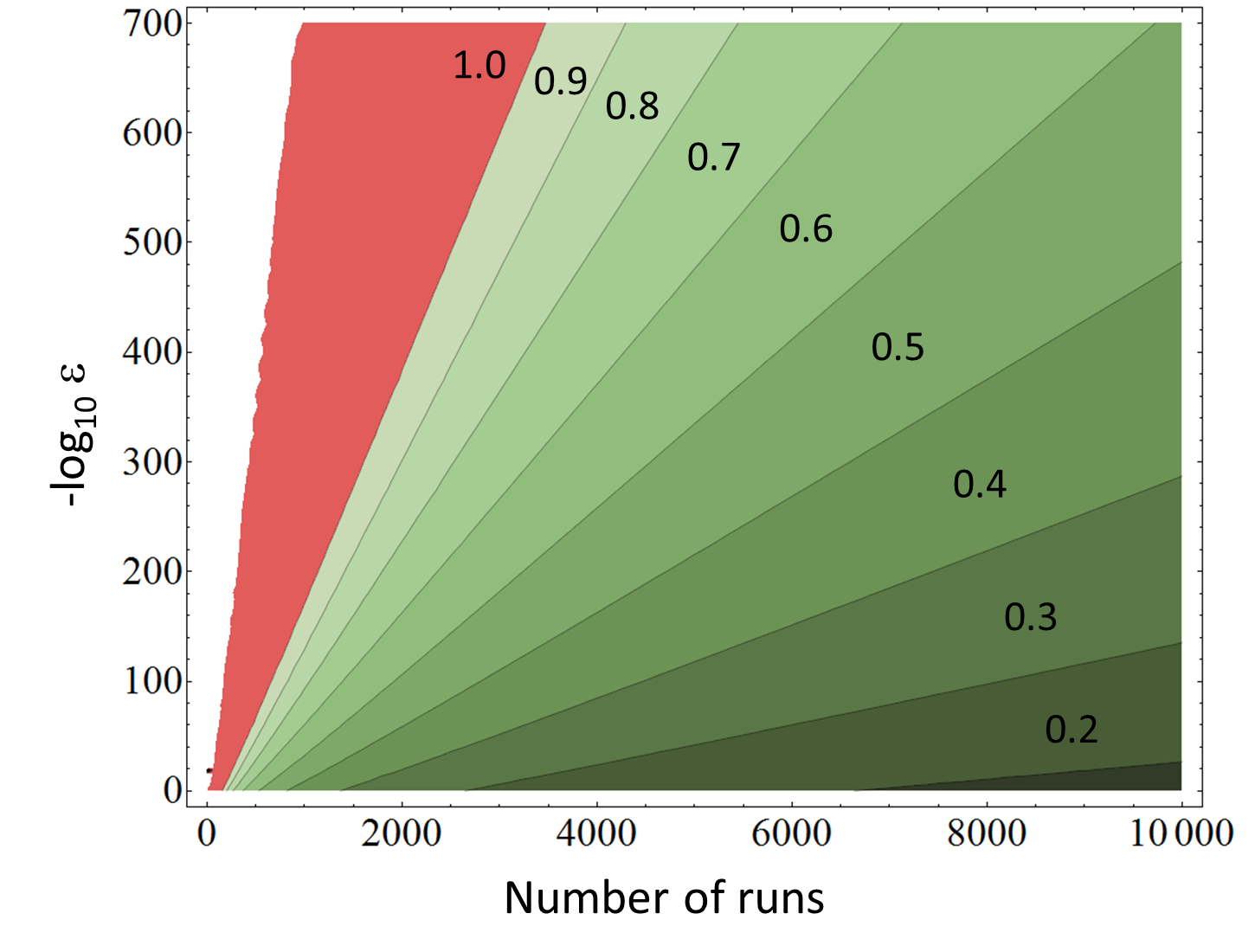}
\caption{(Color online) Contour plot of $\delta$ as a function of the confidence and number of runs $n$. The red region to the farmost left represents the case when $\delta>1$, illustrating a lower bound on the number of runs that must be performed to achieve a certain value of $\epsilon$, quantified by the quantity $-\log_{10} \epsilon$. In practice, even larger values of $n$ will be required to meet a desired confidence.}\label{deltafig}
\end{figure}

On the other hand, the size of the complement region $\overline{\Gamma_i}$ increases as $\delta$ grows larger, implying that large values of $\delta$ result in larger values of $\epsilon_2$. In particular, whenever $\delta\geq 1$ (which can occur for sufficiently low $n$) it will hold that the region $\overline{\Gamma_i}$ will be equal to the entire state space $\Den$ and consequently $\epsilon_2=1$. Thus, for a fixed confidence level, the value of $n$ for which $\delta=1$ sets a lower limit on the number of runs of the experiment that are required to verify the presence of entanglement. This is illustrated in Fig. \ref{deltafig}. These features indicate that in this framework, it is usually necessary to accumulate large amounts of data in order to reliably report the presence of entanglement. 

We have in hand a method to verify whether a set of prescribed regions are in fact confidence regions. The question then remains of how to choose these regions in the first place, an issue that is not addressed in Ref. \cite{ReliableTomo}. Although the results of Christandl and Renner were originally targeted at quantum state tomography, we will instead apply these results in the context of entanglement verification. We now describe a procedure for entanglement verification that fully specifies how to assign confidence regions in terms of entanglement witnesses.


\section{Entanglement verification procedure} 

The goal of an entanglement verification experiment is to determine whether a prepared state is entangled or not with the highest possible certainty. In the language of confidence regions this translates to the task of deciding with the highest level of confidence possible whether the prepared state lies in a region consisting only of entangled states. Reconstructing the state of a general quantum system is experimentally demanding, as the number of required measurement settings will in general increase exponentially with the number of qubits \cite{guehne09a}. Moreover, even if a given state is completely specified, deciding conclusively whether it is entangled is computationally demanding and it is in fact an NP-hard problem in terms of the dimension of the system \cite{gurvits03a}. A way to circumvent these issues is to focus on entanglement witnesses, the use of which has become an increasingly popular tool both in theory and experiments \cite{bruss_EWs_2002,horodecki96a,terhal00a,lewenstein_optimization_2000,bourennane04a}, thus playing a central role in the field of entanglement verification.

A linear entanglement witness $W$ is an observable satisfying
\begin{align*}
&w(\sigma_s):=\tr{\sigma_s W}\geq 0 \hspace{0.2cm} \text{for all }\sigma_s \text{ separable},\\
&w(\sigma_e)<0 \hspace{0.2cm}  \text{for at least one entangled state }\sigma_e.
\end{align*}
Therefore, recording a negative expectation value is a conclusive indicator that the state must have been entangled. We refer to this as the state being \textit{detected} by the witness. Calculating the expectation value of a witness operator can be performed efficiently for any state of arbitrary dimension. Moreover, experimentally determining the expectation value of a witness generally requires considerably fewer measurement settings than a full reconstruction of the state, making them very attractive in practical scenarios. 

One can also consider nonlinear entanglement witnesses \cite{guehne_nonlinear_2006,moroder_iterations_2008} which can be viewed as powerful extensions of linear witnesses in the sense that they will always detect more states than their linear counterparts. Nonlinear witnesses are described by their values $w(\sigma)$ which are nonlinear in the expectation value of the measured observables. They also satisfy the property that their value is negative only for entangled states. Moreover, accessible nonlinear witnesses were recently developed in \cite{NLW}, demonstrating that their expectation value can be evaluated from the same data as the original linear witness. Such nonlinear witnesses have also been recently applied in experiments \cite{EVnonlinear}. 
\begin{figure}
\includegraphics[width=0.69\columnwidth]{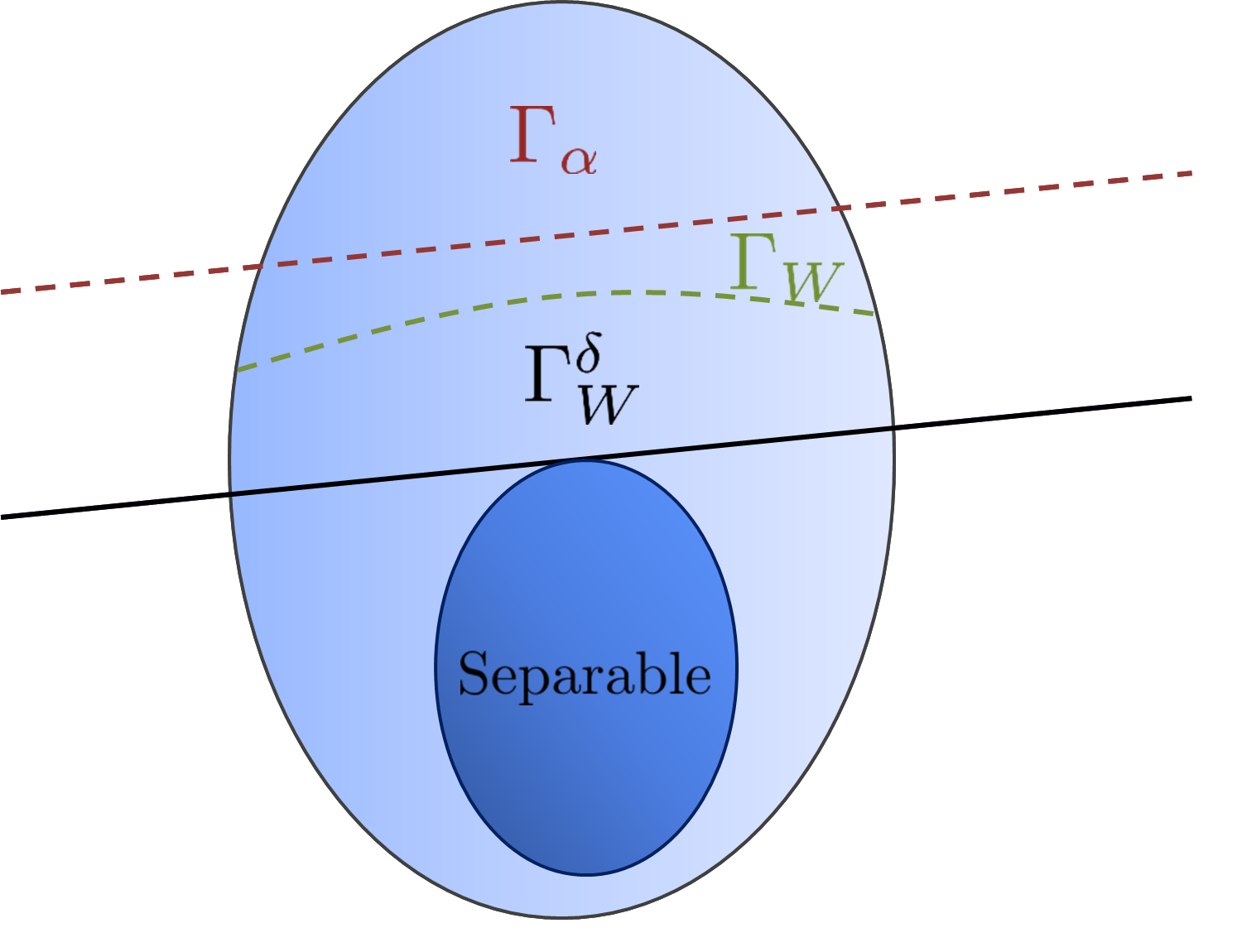}
\caption{(Colour online) The region $\Gd_W$ is fixed as the set of states detected by a linear entanglement witness $W$. This region can be seen as the set of states \textit{above} the black line. Fixing $\Gd_W$ implicitly defines a region $\Gamma_W$ that determines if criteria \eqref{criteria2} is satisfied. This region is located above the dashed green line labelled $\Gamma_W$. The required numerical efforts are greatly simplified by realizing that the set of states $\Gamma_{\alpha}$ above the dashed red line constitute a subset of $\Gamma_W$ as in Observation \ref{observation}.}\label{regions}
\end{figure}

The starting point of our procedure is the specification of an entanglement witness $W$ and a POVM $\{B_i\}$ whose possible outcomes are sufficient to determine the expectation value of $W$. In our description $w(\sigma)$ refers to the value of a linear or nonlinear witness. Recall that in order to verify entanglement whenever it is present, we need to assign confidence regions that contain only entangled states. For this purpose, we define
\begin{align}\label{GammaWDelta}
\Gd_W:=\{\sigma:w(\sigma)<0\}\hspace{0.3cm}\forall i
\end{align}
as the set of detected states. From the definition of an entanglement witness, $\Gd_W$ contains only entangled states. Our goal will be to report $\Gd_W$ as the confidence region whenever possible. Going back to definition \eqref{GammaDelta}, notice that the set $\Gd_i$ is defined for a fixed $\Gamma_i$. But in our picture, we are interested in always reporting regions that contain only entangled states. Therefore, we alternatively choose to fix the reported region $\Gd_W$ and construct the smaller regions implicitly. From \eqref{GammaDelta}, it can be directly seen that if $\Gd_W$ is fixed, its corresponding subregion $\Gamma_W$ is defined by
\begin{align}\label{GammaW}
\Gamma_W:=\{\sigma : \max_{\sigma'\in\overline{\Gd_W}}F(\sigma,\sigma')< \sqrt{1-\delta^2}\}.
\end{align}
We are now ready to specify the mapping from outcomes to regions that constitutes the data analysis procedure for reliable entanglement verification.\\

\textbf{Data analysis procedure.} To construct confidence regions with confidence level $1-\epsilon$ in an entanglement verification experiment, apply the following rule to assign a region to each outcome $B_i$:
\begin{enumerate}
\item{Fix $\epsilon$.}
\item{For each possible measurement outcome $B_i$, compute $\epsilon_2(B_i,\Gamma_W)=\int_{\overline{\Gamma_W}}\mu_i(\sigma) d\sigma$}.
\item{If condition \eqref{criteria2} holds, i.e. if $\epsilon_2\cdot c_{n,d}\leq \epsilon$, assign the set of detected states $\Gd_W$. Otherwise, assign the entire state space $\Den$.}
\end{enumerate}
Therefore, we assign only two possible regions: the set of detected states $\Gamma_W$ or the entire state space $\mathcal{D(H)}$. Note that the entire state space is trivially a confidence region for any given confidence level, so that our assignment indeed produces confidence regions. However, assigning the entire state space must be interpreted as the statement that for the given confidence level, it is not possible to certify that the set of detected states contains the true state. 

Even though the procedure is now completely specified, we are still faced with the difficulty of calculating $\epsilon_2$. As a first step, we note that it is preferable to find a simpler way to characterize the set $\Gamma_W$. One way to do this is to find a subset of $\Gamma_W$ that can be more easily described. We now show that such a subset can always be found in terms of a bound on the expectation value of a linear entanglement witness.

\begin{obs}\label{observation}
Let $W$ be an entanglement witness and let the number $\alpha>0$ satisfy $\alpha>2||W||_{\infty}\delta$. Then the set $\Gamma_{\alpha}=\{\sigma:
\tr{\sigma W}<-\alpha\}$ is a subset of $\Gamma_W$.
\end{obs}

\textit{Proof:} In order to prove the claim we only need to show that $F^2(\sigma,\sigma')< 1-\delta^2$ whenever $\tr{\sigma W}<-\alpha$ and $\tr{\sigma' W}>0$. We begin by considering the following general inequality:
\begin{align}
|\Tr{(\sigma'-\sigma)W}|&=|\langle W,\sigma'-\sigma\rangle|\nonumber\\
&\leq ||W||_{\infty}||\sigma'-\sigma||_{\text{tr}}\nonumber\\
&\leq 2||W||_{\infty}\sqrt{1-F^2(\sigma,\sigma')}\label{fidbound}
\end{align}
where we have used \textit{H\"older's inequality}
\begin{align}
|\langle \sigma,W\rangle|\leq||\sigma||_{\text{tr}}||W||_{\infty}
\end{align} 
and the \textit{Fuchs-van de Graaf inequality} \cite{fuchs1999cryptographic} 
\begin{align}
||\sigma'-\sigma||_{\text{tr}}\leq 2\sqrt{1-F^2(\sigma,\sigma')}.
\end{align}
Now let $\tr{\sigma W}=-\alpha$ and $\tr{\sigma' W}=\beta$ for some $\alpha,\beta>0$. Inserting into \eqref{fidbound} and rearranging we get
\beq
F^2(\sigma,\sigma')\leq 1-\left(\frac{\beta+\alpha}{2||W||_{\infty}}\right)^2\nonumber.
\eeq
We want to find a condition on $\alpha$ such that $F^2(\sigma,\sigma')<1-\delta^2$ for any $\beta$. This will occur whenever
\begin{align*}
1-\left(\frac{\beta+\alpha}{2||W||_{\infty}}\right)^2&<1-\delta^2\\
\Rightarrow \alpha>2||W||_{\infty}&\delta-\beta.
\end{align*}
Since this inequality must hold for all $\beta$, we can restrict ourselves to the worst case scenario of $\beta=0$ to obtain
\beq
\alpha>2||W||_{\infty}\delta
\eeq
as desired.\qed\\

This result is illustrated in Fig. \ref{regions}.  Unfortunately, obtaining a similar and useful result for nonlinear witnesses is difficult: the value of the nonlinear witness may differ greatly for two states even if their fidelity is high. 

Note that because $\Gamma_{\alpha}\subseteq\Gamma_W$, it holds that
\beq
\int_{\overline{\Gamma_{\alpha}}}\Lik d\sigma\geq\int_{\overline{\Gamma_W}}\Lik d\sigma \hspace{0.3cm}\forall i,
\eeq
since $\Lik\geq0$. Therefore if condition \eqref{criteria2} is satisfied when integrating over $\overline{\Gamma_{\alpha}}$, it will always be satisfied for the integral over $\overline{\Gamma_W}$.

Typically, it is possible to assign the set of detected states as a confidence region for very high confidence levels i.e. with $\epsilon\ll 1$. Therefore, from now on we will quantify the confidence level of the procedure by the more appropriate quantity
\begin{align}\label{confidence}
C=-\log_{10}\epsilon,
\end{align}
which we refer to as the \textit{confidence} of the entanglement verification procedure. We further define this quantity to be zero whenever the assigned region is the entire state space $\Den$. Thus, higher values of the confidence result in higher certainty that the state is contained in the set of detected states.

From the description of the data analysis procedure, it should be clear that the crucial step is the computation of $\epsilon_2$: a highly non-trivial task that requires the normalization of the likelihood function as well as its integral over the implicitly defined set $\Gamma_W$. In the following section, we construct and illustrate a series of tools developed to numerically evaluate an upper bound on $\epsilon_2$, ensuring a method to verify condition \eqref{criteria2}.  

\section{Numerical Tools}
There are several difficulties in calculating $\epsilon_2$. An analytical approach is essentially intractable owing primarily to the high dimensionality of parameter space and the non-trivial geometry of the space of positive semi-definite operators \cite{bengtsson06a}. Moreover, the region of integration $\overline{\Gamma_W}$ is not known in closed form but can only be cast as a black-box i.e. we can only ask whether a state lies in this region or not. Finally, we require any approximation of $\epsilon_2$ to provide an upper bound on its value in order to ensure that the inequality $\epsilon_2\leq c_{n,d} \epsilon$ is always satisfied. 

Fortunately, high-dimensional integration over black-box constraints can be handled with the use of Monte Carlo techniques. Most of these techniques are well summarized in \cite{MCMethods}. In the Monte Carlo approach, the mean value of the integrand is approximated by the average value of samples randomly drawn from the region of integration, which in conjunction with knowledge of the hyper-volume of the integration region can be used to calculate the value of the integral. Importantly, the number of samples can be chosen independently of the underlying dimension and any constraint can be straightforwardly incorporated by checking whether a sample point lies within the constraint region.

More specifically, the simplest version of a Monte Carlo technique to approximate a general integral of the form $\int_R f(\sigma)d\sigma$ involves a random sequence of $N$ density operators $\{\sigma_1,\sigma_2,\ldots,\sigma_N\}$ uniformly sampled inside $R$ according to the measure $d\sigma$. By definition, the average $\mean{f}_R$ of a function over a region $R$ satisfies
\beq\label{average}
\int_Rf(\sigma)d\sigma=\mean{f}_R\cdot V_R
\eeq
where $V_R=\int_R d\sigma$ is the hyper-volume of the integration region. The goal in Monte Carlo integration is to approximate the average of the function from the random sample. Namely, we approximate the value of the integral as
\beq
\int_Rf(\sigma)d\sigma\approx \left[\frac{1}{N}\sum_{j=1}^N f(\sigma_j)\right]\cdot V_R,
\eeq
while keeping in mind that all sampled states lie in the integration region. Convergence to the true value of the integral is guaranteed as $N\rightarrow\infty$ due to the law of large numbers \cite{MCMethods}. A main drawback of this approach is that convergence can be extremely slow for highly-peaked functions such as $\mathcal{L}_i(\sigma)$, since only very rarely will a state be drawn from the region surrounding the maximum of the function. This is particularly troublesome for our purposes because an error in the calculation of $\epsilon_2$ can lead to wrong conclusions about the confidence of the procedure. For this reason, we now introduce an approach that can be easily and efficiently implemented and provides an upper bound on $\epsilon_2$. 

We first note that such a bound can be achieved by introducing a lower bound on the normalization constant $\mathcal{N}$. Since the likelihood function is strictly positive, this can always be achieved by integrating over a subset $R$ of $\Den$, i.e.
\beq
\epsilon_2\leq\frac{\int_{\overline{\Gamma_W}}\mathcal{L}_i(\sigma)d\sigma}{\int_{R}\mathcal{L}_i(\sigma)d\sigma}.
\eeq 
We can use this fact to our advantage by restricting $R$ to be a region around the maximum of $\Lik$. Note that this maximum is unique and is in general achieved for a convex set of states \cite{BKEntanglement}. Ideally, this region should be chosen to satisfy $\int_R\Lik d\sigma\approx \int_{\Den}\Lik d\sigma$ in order to provide a tight bound, but this is not necessary as the bound is guaranteed to hold for any $R$. Additionally, because the likelihood function is more flat around the maximum and $R$ is much smaller than $\Den$, drawing random states within $R$ will greatly improve the convergence of a Monte Carlo integration.

We now illustrate how this region $R$ can be constructed from a hyper-rectangle in parameter space. Following the convention of \cite{volmixedstates}, we begin by parametrizing any state $\sigma\in\Den$ in terms of the real-valued Bloch vector $\mathbf{\tau}=(\tau_1,\tau_2,\ldots,\tau_{d^2-1})$ as
\beq
\sigma(\mathbf{\tau})=\frac{1}{d}\id+\sum_{j=1}^{d^2-1}\tau_j \hat{\lambda_j},
\eeq 
where the operators $\{\hat{\lambda_j}\}$ are an orthogonal set of traceless Hermitian generators of $SU(d)$ satisfying $\tr{\hat{\lambda_j}^2}=1$. Any operator written in such a form is immediately Hermitian and of unit trace but may be non-positive for some vectors $\mathbf{\tau}$. Thus, it will be important to keep in mind that not all possible vectors yield valid density matrices. With this parametrization the likelihood function will be a function of the Bloch vector $\Lik=\mathcal{L}_i(\tau_1,\tau_2,\ldots,\tau_{d^2-1})$. Our goal will be to define a region around the maximum that contains only valid states for which the value of the likelihood function is sufficiently large.\\

\textbf{Construction of integration regions.} To construct a region $R$ to be used in an approximation of the normalization of the likelihood function, perform the following:
\begin{enumerate}
\item{Calculate the maximum value of the likelihood function $\mathcal{L}_i^{\text{max}}$ and any vector $\mathbf{\tau}^*=(\tau_1^*,\tau_2^*,\ldots,\tau_{d^2-1}^*)$ for which this maximum is attained.}
\item{Find, for all $j$, the lowest possible quantities $x_j^{\pm}>0$ such that $\mathcal{L}_i(\tau_1^*,\tau_2^*,\ldots,\tau_j^*\pm x_j^{\pm},\ldots, \tau_{d^2-1}^*)=\mathcal{L}_i^{\text{max}}/\eta$ for some fixed number $\eta>0$. If no such values can be found for some $j$, let $x_j^{\pm}=\infty$. }
\item{Find, for all $j$, the highest possible quantities $y_j^{\pm}>0$ such that $\sigma(\tau_1^*,\tau_2^*,\ldots,\tau_j^*\pm y_j^{\pm},\ldots, \tau_{d^2-1}^*)$ is still a valid density matrix.}
\item{Define $r_j^{\pm}=\min\{x_j^{\pm},y_j^{\pm} \}$. Then the integration region $R$ is equal to all the valid density matrices within the hyper-rectangle $r$ defined by $r=\{\mathbf{\tau}:\tau_j^*-r_j^-\leq \tau_j\leq \tau_j^*+r_j^+,\forall j\}$.}
\end{enumerate}

\begin{figure}
\includegraphics[width=0.79\columnwidth]{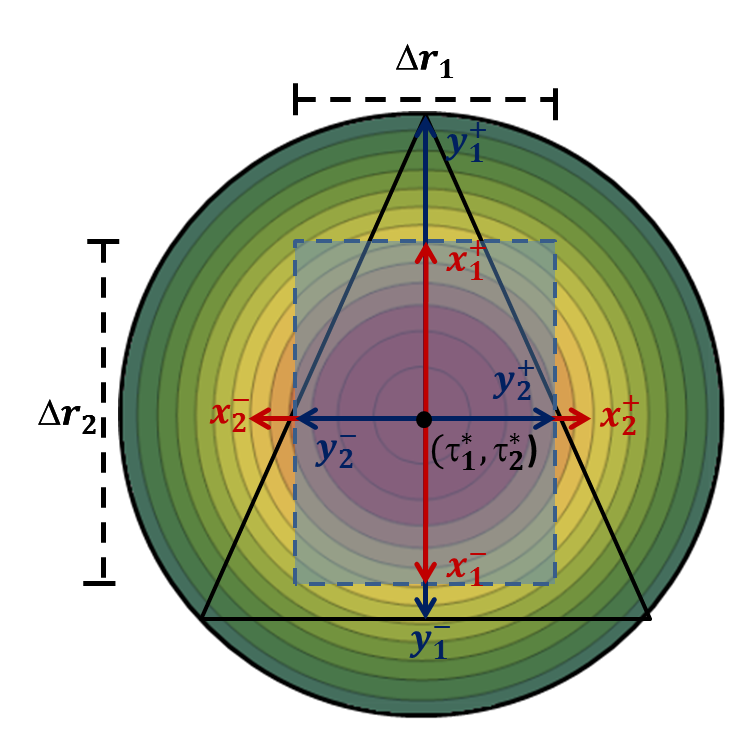}
\caption{(Color online) Construction of integration regions. We imagine a two-dimensional section of parameter space characterized by the variables $\tau_1$ and $\tau_2$. Only the region inside the triangle contains valid density matrices and contours of $\Lik$ are shown in the background. To construct the integration region we do the following: 1. Find the maximum of the function and a state for which it occurs, in this case $(\tau_1^*,\tau_2^*)$. 2. From this maximum, find the displacements $x_1^{\pm}$  and $x_2^{\pm}$ such that the value of the function is decreased by a specified amount, in this case corresponding to the 6th contour line. 3. Find the displacements $y_1^{\pm}$  and $y_2^{\pm}$ that define the points where the boundary of valid states is met. 4. By choosing the minimum of these quantities in each direction, we construct a rectangle (dashed) and the integration region is the intersection of this rectangle with the space of valid density matrices.}\label{integration}
\end{figure}

This construction is illustrated in Fig. \ref{integration}. Note that the task of maximizing the likelihood function can be performed efficiently and is routine in the context of quantum state tomography. A good choice of $\eta$ will in general depend on each individual problem, but it should be chosen large enough to include only regions that contribute significantly to the integral.

Once the hyper-rectangle has been constructed, it is straightforward to perform the Monte Carlo integration by sampling uniformly within the rectangle, while keeping only operators in that sample that are valid density matrices. Let these sampled states form the set $\{\sigma_1,\sigma_2,\ldots,\sigma_N\}$. The target integral is then given by
\begin{align}
\int_R \Lik d\sigma&\approx\left[\frac{1}{N}\sum_{j=1}^N \mathcal{L}_i(\sigma_j)\right]\cdot V_R\nonumber\\
&=\mean{\mathcal{L}_i}_R\cdot V_R\label{targetint}.
\end{align}
Because typical values of the likelihood function are extremely small, it is preferable to work with the logarithm of the function and use the identity
\begin{align}
\log{(a+b)}=\log[\exp(\log a-\log b)+1 ]+\log b
\end{align}
to add the values of $\mathcal{L}_i(\sigma_j)$ at each step of the algorithm and determine $\mean{\mathcal{L}_i}_R$ as in equation \eqref{targetint}.

In order to calculate $V_R$, we use the fact that the Hilbert-Schmidt metric on the space of quantum states generates the Hilbert-Schmidt measure \cite{volmixedstates}. The Hilbert-Schmidt distance between two density matrices is given by
\beq
D_{HS}(\sigma_1,\sigma_2)=||\sigma_1-\sigma_2||_2=\sqrt{\Tr{(\sigma_1-\sigma_2)^2}}.
\eeq
This correspondence between metric and measure implies that the volume of the hyper-rectangle $r$ can be found in the usual sense as the product of the length of its sides with respect to the Hilbert-Schmidt metric.  More specifically, let $\sigma_j^{\pm}=\sigma(\tau^*_1,\ldots,\tau^*_j\pm r_j^{\pm},\ldots,\tau^*_{d^2-1})$. Then the length $\Delta r_j$ of the $j$th side of $r$ is given simply by
\begin{align}
&\Delta r_j=D_{HS}(\sigma_j^+,\sigma_j^-)\nonumber\\
&=||r_j^+\hat{\lambda_j}+r_j^-\hat{\lambda_j}||_2=r_j^++r_j^-,
\end{align}
where we have used the fact that the operators $\hat{\lambda_j}$ are normalized with respect to the Hilbert-Schmidt inner product. The hyper-volume $V_r$ of $r$ is then given by
\beq
V_r=\prod_{j=1}^{d^2-1}\Delta r_j.
\eeq

This correspondence is also useful in generating a random sample, as one needs only to obtain a random number within the intervals $[\tau_j^*-r_j^-,\tau_j^*+r_j^+]$. Because not all operators in $r$ are valid density matrices, $V_r$ is in general larger than the hyper-volume $V_R$ of the integration region $R$. However, one can estimate $R$ from knowledge of the fraction $f$ of the randomly drawn operators that are valid density matrices. The relationship between these quantities is
\beq
V_R\approx f\cdot\prod_{j=1}^{d^2-1}\Delta r_j,
\eeq
which can finally be inserted in \eqref{targetint} to provide the numerical calculation of the target integral
\beq
\int_R \Lik d\sigma\approx\left[\frac{1}{N}\sum_{j=1}^N \mathcal{L}_i(\sigma_j)\right]\cdot f\cdot\prod_{k=1}^{d^2-1}\Delta r_k.
\eeq

To calculate $f$, it is sufficient to verify how many of the drawn operators are valid density operators and divide this number by the total number of randomly drawn operators.

One could imagine that a similar technique could be used to calculate the integral $\int_{\overline{\Gamma_W}}\Lik d\sigma$  appearing in the definition of $\epsilon_2$. Unfortunately, this would greatly increase the computational efforts in the construction of $R$ since one must additionally ensure that each of the drawn samples lie in $\Gamma_W$. Additionally, in this case restricting the integration region results in an incorrect lower bound on $\epsilon_2$. Instead, we can construct an upper bound on this integral via the maximum of the likelihood function as
\begin{align}
\int_{\overline{\Gamma_W}}\Lik d\sigma&=\mean{\mathcal{L}_i}_{\overline{\Gamma_W}}\cdot V_{\overline{\Gamma_W}}\nonumber\\
&\leq\left(\max_{\sigma\in{\overline{\Gamma_W}}}\Lik\right)\cdot V_{\Den},
\end{align}
where $V_{\Den}$ is the Hilbert-Schmidt hyper-volume of the entire state space. This volume was calculated explicitly in \cite{volmixedstates} for Hilbert spaces of arbitrary dimension. We can then combine this result with our previous bound on the normalization constant to provide an overall upper bound on $\epsilon_2$. Since this value will be typically very small and in order to make a direct comparison with the confidence, we will henceforth refer to the logarithm of $\epsilon_2$ for which we now have the inequality
\beq
\log_{10}\epsilon_2\leq\log_{10}\left(\frac{\max_{\sigma\in{\overline{\Gamma_W}}}\Lik}{\mean{\mathcal{L}_i}_R}\frac{V_{\Den}}{V_R}\right).
\eeq 

Of course, the average of the likelihood function over $\overline{\Gamma_W}$ will generally be much smaller than the maximum over this region, making the bound very loose. However, in practice this is not a problem because the above bound on $\epsilon_2$ is dominated by the much larger differences between the global maximum of the function and its maximum over $\overline{\Gamma_W}$. More specifically, for experiments with a large number of runs (large $n$), it will typically hold that
\beq
|\log_{10}\left(\frac{\max_{\sigma\in{\overline{\Gamma_W}}}\Lik}{\mean{\mathcal{L}_i}_R}\right)|\gg|\log_{10}\left(\frac{\mean{\mathcal{L}_i}_{\overline{\Gamma_W}}}{\max_{\sigma\in{\overline{\Gamma_W}}}\Lik}\right)|,
\eeq 
so that 
\begin{align}
&\log_{10}\left(\frac{\mean{\mathcal{L}_i}_{\overline{\Gamma_W}}}{\mean{\mathcal{L}_i}_R}\right)=\nonumber\\
&\log_{10}\left(\frac{\max_{\sigma\in\overline{\Gamma_W}}\Lik}{\mean{\mathcal{L}_i}_R}\frac{\mean{\mathcal{L}_i}_{\overline{\Gamma_W}}}{\max_{\sigma\in{\overline{\Gamma_W}}}\Lik}\right)\nonumber\\
&\approx \frac{\max_{\sigma\in{\overline{\Gamma_W}}}\Lik}{\mean{\mathcal{L}_i}_R}
\end{align}
and the value for $\log_{10}\epsilon_2$ is not altered significantly by the loose bound.

The final quantity we must be able to calculate is the maximum of the likelihood function over $\overline{\Gamma_W}$. This again is a non-trivial global optimization problem involving a black-box constraint. As in the case of integration, the particular features of this problem impede the usual techniques and strongly ask for a Monte Carlo approach. To handle the optimization in the general case, we employ an adaptation to the quantum scenario of a simulated annealing algorithm (SA)  
based on the Metropolis-Hastings algorithm outlined in \cite{BKMetropolis}.

The SA algorithm is based on a biased random walk that preferentially moves to states with higher values of the objective function while still accepting moves to lower values with a probability governed by a global ``temperature'' parameter. This last feature prevents the algorithm from being confined in local maxima. Unfortunately, this same feature makes the convergence slow, usually requiring many steps to reach close proximity to the maximum. For each step, one must additionally make the costly verification that the states lie in the region of integration $\overline{\Gamma_W}$, so it must be understood that run times are usually long.  A detailed description of the algorithm is included in the Appendix.

One drawback of the SA algorithm is that due to its stochastic nature, independent runs of the algorithm will generally yield different values. Moreover, by construction these values cannot be larger than the global maximum. In order to address this issue, one should estimate the numerical error by performing many independent runs of the algorithm and collecting statistics of the sample values. The usual choice is to calculate the standard deviation of the values \cite{MCMethods} and take this as the error. It is then important to ensure that condition \eqref{criteria2} is satisfied well within this error.

Nevertheless, we are still interested in obtaining a more efficient method to solve the maximization problem. We can achieve this for the case of linear witnesses by noting that for the subset $\Gamma_{\alpha}$ of $\Gamma_W$, it holds that
\beq
\max_{\sigma\in\overline{\Gamma_{\alpha}}}\Lik\geq\max_{\sigma\in\overline{\Gamma_{W}}}\Lik
\eeq
since in that case $\overline{\Gamma_W}$ is a subset of $\overline{\Gamma_{\alpha}}$. Therefore, we can provide a final expression for the bound on $\epsilon_2$ as
\beq
\log_{10}\epsilon_2\leq\log_{10}\left(\frac{\max_{\sigma\in{\overline{\Gamma_{\alpha}}}}\Lik}{\mean{\mathcal{L}_i}_R}\frac{V_{\Den}}{V_R}\right)
\eeq
where $\Gamma_{\alpha}$ is defined as in Observation \ref{observation}. This expression has the enormous advantage that because the constraint over $\Gamma_{\alpha}$ is convex and $\Lik$ is log-convex, the maximization of $\Lik$ over this region can be calculated with vastly greater efficiency using standard methods in convex optimization.

We are additionally interested in reporting the highest possible confidence level, which corresponds to the case in which the equality $\epsilon_2\cdot c_{n,d}=\epsilon$ holds. The value of $\epsilon_2$ depends on the region $\Gamma_W$ which in turn implicitly depends on $\epsilon$ through the definition of the enlarging parameter $\delta$, so that the above equality is in principle an equation to be solved for $\epsilon$. Unfortunately, there is no clear method of how to solve the equation directly, primarily because of the difficulty of calculating $\epsilon_2$ itself. Instead, to achieve the highest possible confidence level, one must iteratively adapt the chosen value of $\epsilon$ until $\epsilon_2\cdot c_{n,d}\approx\epsilon$ while still satisfying the inequality \eqref{criteria2}. 

With these tools in hand it is now possible to apply the reliable entanglement verification procedure for both linear and nonlinear witnesses. We now proceed to demonstrate the features of the method by applying the technique to data obtained from an experiment generating a family of entangled two-photon states. The entanglement of these states is verified with the use of an ANLW.

\section{Experiment} 
\begin{figure}
  \begin{center}
   \includegraphics[width=0.7\columnwidth]{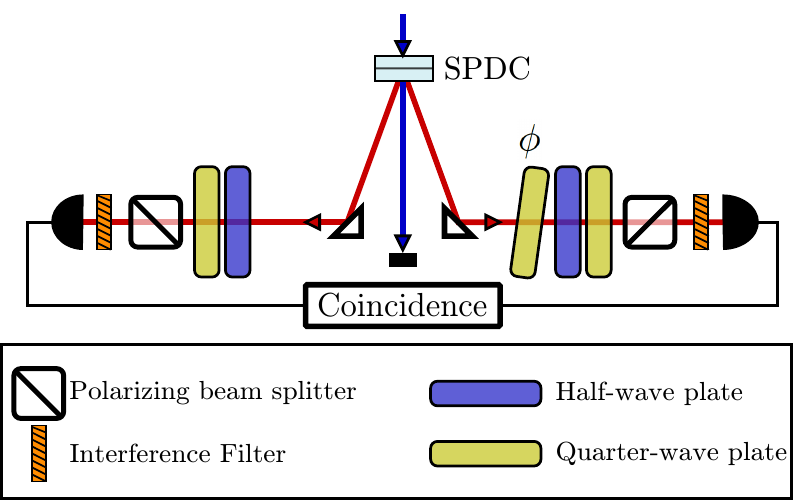}
  \end{center}
 \caption{(Color online.) Experimental setup for producing $\ket{\Phi(\phi)}=\frac{1}{\sqrt{2}}\left(\ket{HH}+e^{i\phi}\ket{VV}\right)$ polarization states. Photon pairs are generated via type-I noncollinear SPDC in a pair of orthogonally oriented BBO crystals and analyzed with wave plates and polarizing beamsplitters. The phase $\phi$ is adjusted by tilting a quarter-wave plate.\label{expsetup}}
\end{figure}

To apply our entanglement verification procedure to experimental data, we aimed to produce photon pairs in the maximally entangled states $\ket{\Phi(\phi)}=\frac{1}{\sqrt{2}}\left(\ket{HH}+e^{i\phi}\ket{VV}\right)$, where $\ket{H}$ and $\ket{V}$ are defined respectively as polarization parallel and perpendicular to the optical table. A frequency doubled titanium-sapphire laser (80~MHz, 790~nm) was used to pump a pair of orthogonally oriented 1~mm $\beta$-Barium borate (BBO) crystals, as seen in Fig.~\ref{expsetup}. By pumping with diagonal polarization $\ket{D}=\frac{1}{\sqrt{2}}\left(\ket{H}+\ket{V}\right)$, the pump may produce photon pairs via type-I noncollinear spontaneous parametric down-conversion (SPDC) in either the first or second crystal~\cite{KwiatSource}. Bismuth borate, $\alpha$-BBO, and quartz crystals were used to ensure that each path was spatially and temporally indistinguishable, and the photon pairs were filtered using bandpass filters with a centre wavelength of 790~nm and a bandwidth FWHM of 3~nm. The single photon signal was measured with avalanche photodiodes (APDs) and coincidences were recorded within a 3~ns window.

Single photons were detected at a rate of approximately 200~kHz in each arm, with a coincidence rate of approximately 35~kHz when the measurements are set to $HH$ or $VV$. A quarter-wave plate was tilted to introduce an arbitrary phase shift between horizontally and vertically polarized components, allowing control over the phase $\phi$. This setup constitutes part of the setup used for the experiment reported in \cite{lavoie_experimental_2010}. The two-photon state was prepared for six values of $\phi$, corresponding to a waveplate tilt range of twelve degrees and transforming the state from $\ket{\Phi^-}$ to $\ket{\Phi^+}$.

Projective measurements were taken in three bases, corresponding to the eigenbases of the operators $\{\sx\otimes\sx,\sy\otimes\sy,\sz\otimes\sz\}$. We will refer to the elements of these bases as $\ketbra{x_i}{x_i}$, $\ketbra{y_i}{y_i}$ and $\ketbra{z_i}{z_i}$ respectively. For example, the eigenbasis of $\sz\otimes\sz$ is given by $\ket{z_1}=\ket{HH},\ket{z_2}=\ket{HV},\ket{z_3}=\ket{VH},\ket{z_4}=\ket{VV}$, and similarly for the other bases. To verify the entanglement of these states, an accessible nonlinear witness was constructed from the linear witness $W=(1/4)(\id+\sx\otimes\sx-\sy\otimes\sy+\sz\otimes\sz)$. Following \cite{NLW}, the expectation value $w_{\infty}(\sigma)$ of the nonlinear witness for a state $\sigma$ can be expressed as
\begin{align}
w_{\infty}(\sigma)=\tr{\rho W}-|c|^2-\frac{|d|^2}{1-|k|^2},
\end{align}
where
\begin{align}
c&=\Tr{\sigma (\ketbra{\psi^-}{\psi^-}U)^{t}}\nonumber\\
k&=\tr{\sigma U^{t}}\nonumber\\
d&=\tr{\sigma W}-ck,\nonumber
\end{align}
$\ket{\psi^-}=\frac{1}{\sqrt{2}}(\ket{HV}-\ket{VH})$ and the superscript $t$ denotes partial transposition. By choosing $U=\sz\otimes\sz$, this expectation value can be computed from the expectation value of the aforementioned operators and the nonlinear witness is \textit{accessible} \cite{NLW}. An accessible nonlinear witness was chosen because it detects these entangled states for most values of $\phi$.

\begin{figure}
  \begin{center}
   \includegraphics[width=\columnwidth]{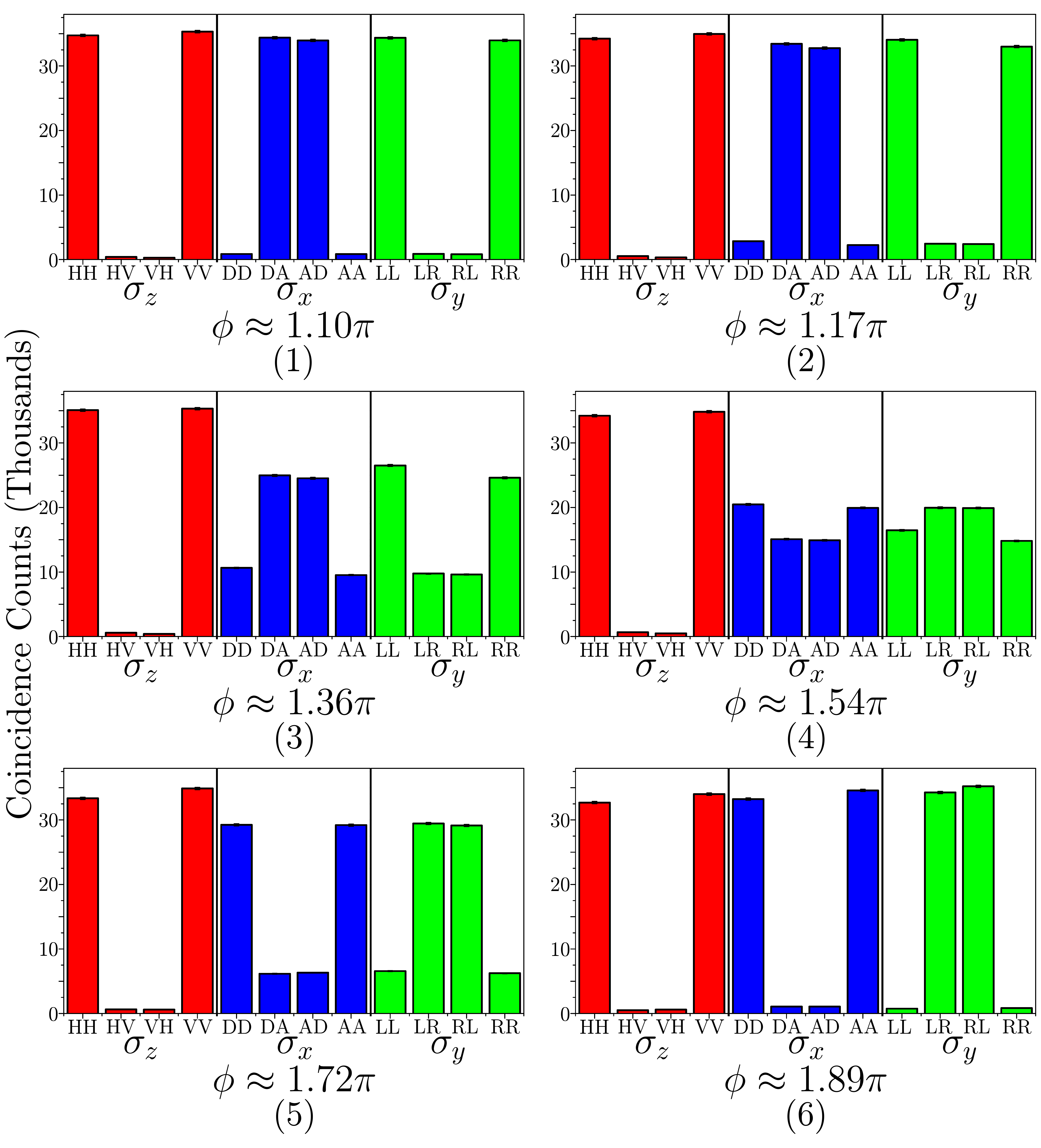}
  \end{center}
 \caption{(Color online.) Results of projective measurements on six states of the form $\frac{1}{\sqrt{2}}\left(\ket{HH}+e^{i\phi}\ket{VV}\right)$, corresponding to the eigenbases of the operators $\{\sx\otimes\sx,\sy\otimes\sy,\sz\otimes\sz\}$. The approximate value of the phase is included for each case. Counts were integrated over 1 s per measurement setting.\label{bars}}
\end{figure}

In this experiment, all measurements are independent so that each element of the POVM $\{B_i\}$ is a tensor product of the operators corresponding to possible individual outcomes. The likelihood function takes the form
\begin{align}
\Lik=&\prod_{j=1}^4\tr{\sigma\ketbra{x_j}{x_j}}^{n_x^j}\cdot\tr{\sigma\ketbra{y_j}{y_j}}^{n_y^j}\nonumber\\
&\cdot\tr{\sigma\ketbra{z_j}{z_j}}^{n_z^j},
\end{align}
where $n_x^j$ is the number of times outcome $\ketbra{x_j}{x_j}$ is obtained and similar definitions hold for the other operators, so that the total number of measurement outcomes is $n=\sum_{j=1}^4n_x^j+n_y^j+n_z^j$. Note that in this case the measurement outcome $B_i$ is fully specified by the numbers $\{n_x^j,n_y^j,n_z^j\}$. In the experiment, six states were prepared corresponding to six different values of the parameter $\phi$. The measurement outcomes for each case are summarized in Fig. \ref{bars}.

We have calculated the confidence as in equation \eqref{confidence} for the six preparations of the entire experiment. These results are illustrated in Table \ref{ExpResults}. We can report very high confidences for almost all states, with the exception of state 4 for which condition \eqref{criteria2} cannot be satisfied for any value of $\epsilon$. This is not entirely surprising as this state presents the weakest correlations in the $\{\ketbra{x_j}{x_j}\}$ and $\{\ketbra{y_j}{y_j}\}$ bases leading to a value of the nonlinear witness that is closest to zero, as seen in Table \ref{ExpResults}. Thus, the outcomes for this case most closely resemble the ones that could be obtained from a separable state. This again is evidence that only large data which are clearly inconsistent with separable states can lead to the reliable statements obtained from our procedure.

\begin{table}
\begin{tabular}{|c|c|c|c|}\hline
State & Approximate phase & Confidence & $w_{\infty}$\\\hline
1 & $1.10\pi$ & 5150 & -23.0\\
2 & $1.17\pi$ & 2050 & -15.2\\
3 & $1.36\pi$ & 410 & -3.4\\
4 & $1.54\pi$ & 0 & -0.3\\
5 & $1.72\pi$ & 1819 & -5.8\\
6 & $1.89\pi$ & 4980 & -13.6\\
\hline
\end{tabular}
  \caption{Calculation of the confidence and value of the nonlinear witness for all prepared states in the experiment. The total number of counts obtained in each case was roughly 35,000.}
  \label{ExpResults}
\end{table}

\begin{figure}
  \begin{center}
   \includegraphics[width=0.8\columnwidth]{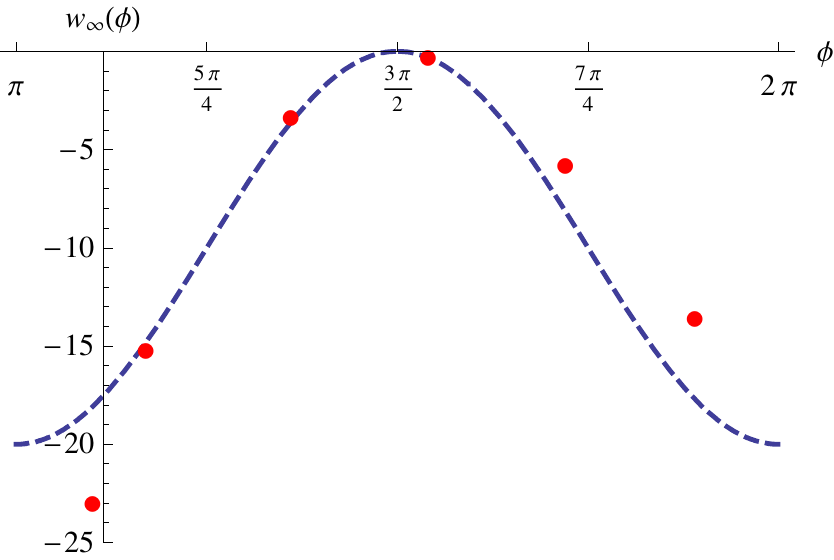}
  \end{center}
 \caption{(Color online) Value of the nonlinear witness $w_{\infty}(\phi)$ for the six states prepared in the experiment (dots). The value of the nonlinear witness for the family of states $\sigma(\phi)=(1-p)\ketbra{\Phi(\phi)}{\Phi(\phi)}+\frac{p}{4}\id$ with $p=1/42$ is shown in the background (dashed). This curve is included only to illustrate the values of $\phi$ for which it is difficult to verify entanglement and should not be interpreted as a fit to the data. The value of $p$ was chosen to adjust the scaling to the recorded values. \label{dataplot}}
\end{figure}

\begin{figure}
  \begin{center}
   \includegraphics[width=\columnwidth]{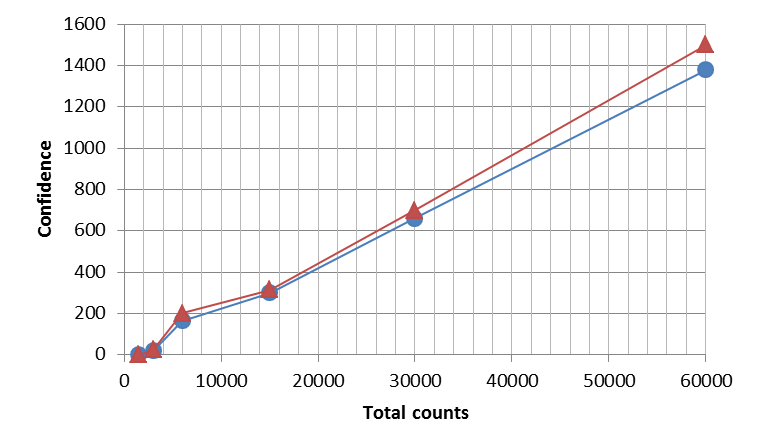}
  \end{center}
 \caption{(Color online) Confidence for random samples of different size, quantified by the total number of counts. The confidences were calculated for $\Gamma_W$ (triangles) and $\Gamma_{\alpha}$ (dots). These results illustrate that the bound introduced by considering the subset $\Gamma_{\alpha}$ is small and is not an impediment to reach a very large confidence. In the case of 1500 total counts, the confidence is zero, consistent with our understanding that a large number of outcomes are needed in order to reliably report entanglement with our technique. Moreover, the data shows that the confidence is roughly linear in the number of outcomes.\label{samples}}
\end{figure}

Additionally, we are interested in understanding how the maximum achievable confidence depends on the total number of runs of an experiment. It is also important to gain insight on the cost of using the bound of Observation \ref{observation} for linear witnesses. For this purpose, samples of different size were randomly selected from the outcomes of experiment (6) in Fig. \ref{bars}. That is, from the entire set of observations in this experiment (shown in Fig. \ref{bars}), we randomly selected a subset of all the data and interpreted it as arising from an experiment with a fewer number of runs (counts). The confidence was calculated for each of them using both regions $\Gamma_W$ and $\Gamma_{\alpha}$, this latter being possible because this state is also detected by the linear witness. The obtained values using these two different methods is portrayed in Fig. \ref{samples} and Table \ref{samplestable}.

\begin{table}[hbt!]
\begin{tabular}{|c|c|c|}\hline
Total counts &Confidence $(\Gamma_{\alpha})$ & Confidence $(\Gamma_W)$\\\hline
1500 & 0 & 0\\
3000 & 18 & 24\\
6000 & 165 & 200\\
15000 & 300 & 315\\
30000 & 660 & 700\\
60000 & 1378 & 1500\\
\hline
\end{tabular}
  \caption{Calculation of the confidence for samples of different size from the outcomes of experiment 6 based on $\Gamma_W$ and $\Gamma_{\alpha}$. \label{samplestable}}
  \label{samplestable}
\end{table}

The results indicate that, as a percentage of the total confidence, the loss introduced by considering $\Gamma_{\alpha}$ is small. It is also clear that a large number of runs are necessary in order to report a non-zero confidence, in accordance to our understanding of the role of the enlarging parameter $\delta$ and the polynomial factor $c_{n,d}$ as discussed in section \ref{confreg}. To estimate the numerical error present in the SA algorithm,  we performed 20 independent runs of the algorithm for the data of state 1 and found this numerical error to be $1.85\%$. In all calculations it was ensured that condition \eqref{criteria2} was satisfied by at least ten times this numerical error. In the construction of the integration regions a value of $\eta=10^5$ was chosen for all cases. Finally, the CVX package for specifying and solving convex programs \cite{cvx} was used to numerically calculate the global maximum of the likelihood function, as well as its maximum over $\Gamma_{\alpha}$. 

\section{Conclusion}
In this paper, we have applied the work of M. Christandl and R. Renner in Ref. \cite{ReliableTomo} to the case of entanglement verification. Through the concept of confidence regions, we have provided a procedure to make reliable and efficient statistical statements quantifying the confidence level of having entanglement present in a physical system. These statements have a clear operational interpretation and do not require the specification of a prior distribution nor the assumption of independent measurements or i.i.d. sources. We have shown that this method can be applied in practice by developing specific numerical tools designed to calculate all necessary quantities. For the particular case of experiments relying on linear entanglement witnesses, we have shown that the procedure can be implemented efficiently using only plain Monte Carlo integration and convex optimization methods. The procedure is ready to be applied to current experiments as we demonstrated by applying the technique to data obtained from an experiment generating entangled two-photon states. High confidence values can be achieved whenever the data is strongly inconsistent with a separable state and the number of measurement outcomes is large enough. Our results thus provide an illustration of the techniques that must be employed in current experiments in order to obtain clear and reliable claims. 

It is important to note that this work assumes that there are no systematic errors in the measurements performed. In any real experiment, there will always be discrepancy between the intended measurement and the one actually performed, no matter how small this discrepancy is. These systematic errors can in principle lead to incorrect statements and a method to incorporate it in the framework must be pursued. Numerical techniques also invariably involve errors and these should also be clearly incorporated in the framework. Future research may lead to improved algorithms. Finally, let us note that it is often desirable to quantify the amount of entanglement present as opposed to just verifying it. Our technique can in principle be applied to such cases by reporting regions that contain states with at least a certain amount of entanglement. Future work can focus on including this feature into the procedure.

\textit{Acknowledgements.-} We thank Chris Ferrie for his insight on numerical methods to integrate probability distributions and Matthias Christandl and Renato Renner for their hospitality in Z\"urich and fruitful discussions concerning their work. Most importantly, we thank Philippe Faist for valuable discussions on his work and on convex optimization techniques. Oleg Gittsovich is grateful for the support of the Austrian Science Fund (FWF) and Marie Curie Actions (Erwin Schr\"odinger Stipendium J3312-N27). This work was supported by Industry Canada, Ontario Ministry of Research and Innovation ERA, QuantumWorks, Ontario Centres of Excellence, the Canadian Foundation for Innovation, and NSERC Discovery and NSERC Strategic Project Grant (SPG) FREQUENCY.

\appendix*
\section*{Appendix}
Here we fully describe the simulated annealing (SA) algorithm. The algorithm is based on a biased random walk in state space that preferentially selects states with a higher value of the likelihood function at each new step of the iteration. However, it also accepts jumps to states with lower values with a probability that depends on a global parameter $T$, usually referred to as the temperature because of its similarity with the physical temperature in the annealing process of metallurgy.

\begin{figure}
  \begin{center}
   \includegraphics[width=\columnwidth]{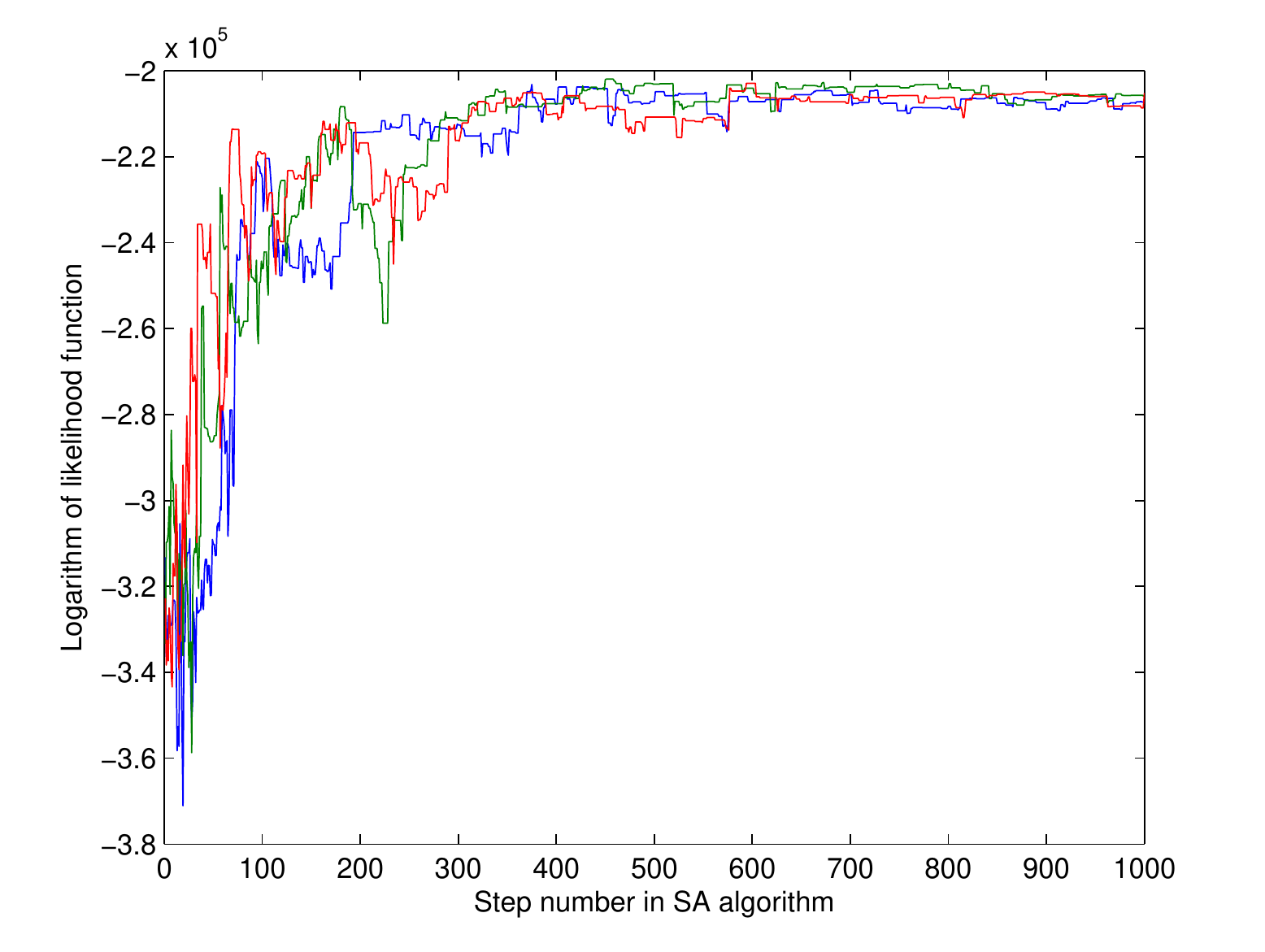}
  \end{center}
 \caption{(Color online) Three independent runs of the same simulated annealing algorithm for the data of experiment 1. Although all parameters are identical in each case, the output is slightly different in each case due to the stochastic nature of the algorithm.\label{SAruns}}
\end{figure}

Below is a full enumeration of all the steps of the algorithm to calculate the maximum value of the likelihood function $\mathcal{L}(\sigma)$ over the set $\Gamma_W$. A graphical illustration of how the maximum value of the function is reached as the algorithm progresses is found in Fig. \ref{SAruns}. The random walk here described is based upon the quantum adaptation of the Metropolis-Hastings algorithm depicted in \cite{BKMetropolis}.\\

\textbf{Simulated annealing algorithm:}

\begin{enumerate}
\item{Select an initial value $T_0$ for the temperature $T$ as well as for the ``step size" $\Delta$.  }
\item{Generate a $d\times d-$dimensional random state $\ket{\psi}$ according to the Haar measure, where $d$ is the dimension of the underlying Hilbert space $\Hil$. Trace out one of the subsystems to obtain the state $\sigma_0$. If $\sigma_0\in\overline{\Gamma_W}$ continue to the next step, repeat otherwise.}
\item{Randomly choose a $2\times 2$ Hermitian matrix $H_{kl}$ in the following way. Pick two integers $k,l$ randomly from the set $\{1,2,\ldots,d\}$. If $k<l\rightarrow H_{kl}= \ketbra{k}{l}+\ketbra{l}{k}$, similarly if $k>l\rightarrow H_{kl}=-i\ketbra{k}{l}+i\ketbra{l}{k}$ and finally if $k=l\rightarrow H_{kl}=\ketbra{k}{k}-\ketbra{k+1}{k+1}$ (set $k+1=1$ if $k=d$).}
\item{Pick a distance $\delta$ by sampling from a Gaussian distribution with mean 0 and standard deviation $\Delta$.}
\item{Compute the state $\ket{\psi'}=\exp(i H_{kl} \delta)\ket{\psi}$. Trace out one of the subsystems of $\ket{\psi'}$ to obtain the state $\sigma'_0$.}
\item{If $\sigma_0'\notin\overline{\Gamma_W}$, repeat steps 2 to 5, continue otherwise.}
\item{Evaluate the ratio $R=\log \left(\mathcal{L}(\sigma'_0)/\mathcal{L}(\sigma_0)\right)$. If $R>0$ $(\mathcal{L}(\sigma'_0)>\mathcal{L}(\sigma_0))$, let $\sigma_1=\sigma'_0$. Otherwise, flip a coin with bias $p=\exp\{-|\log(\mathcal{L}(\sigma_0))-\log(\mathcal{L}(\sigma_0'))|/T\}$.  If ``1" is obtained (which happens with probability $p$), again let $\sigma_1=\sigma'_0$, otherwise $\sigma_1=\sigma_0$.}
\item{Repeat steps 2-6 $N$ times to generate a set $\{\sigma_1,\sigma_2,\ldots,\sigma_N\}$ corresponding to $N$ steps of the random walk. For each step, adapt the temperature via the cooling rule $T(s)=T_0/s$ where $s$ is the step of the walk. The maximum value of $\mathcal{L}(\sigma)$ over this set is the output of the algorithm.}
\end{enumerate}

The performance of the algorithm depends strongly on the value of $\Delta$ and this value must be adapted throughout each step of the walk in order to maintain a fixed average acceptance ratio, i.e. the fraction of times we jump to a new state. Various values for these ratios are suggested \cite{DeltaValues}. Similarly, the choice of initial temperature is crucial. Its role is to prevent the algorithm from being stuck in local maxima by allowing it to escape such cases in the initial stages of the algorithm. The temperature is then reduced to ensure that convergence to the maximum is attained. Therefore, the choice of initial temperature and cooling rule is essential and varies for different cases. In practice, they must be chosen for each particular problem based mostly on experience.

Finally, in order to check whether a new state belongs in $\overline{\Gamma_W}$, it is necessary to determine the maximum fidelity of this state with any state in this set. For this purpose, we exploit the fact that the fidelity function is concave in both its arguments and that the restriction $\rho\in\overline{\Gd_W}$ is convex for both linear and nonlinear witnesses. These properties allow us to employ the highly efficient tools of convex optimization to solve the maximization problem. Concretely, for a given state $\sigma$, we verify membership in $\overline{\Gamma_W}$ by solving the problem
\begin{align}
\text{maximize}\hspace{0.2cm} F(\sigma,\sigma')\nonumber\\
\text{subject to}\hspace{0.2cm} \sigma'\in \overline{\Gd_W}\nonumber
\end{align}
where $\sigma'$ must be forced to be a density operator. The state $\sigma$ is a member of $\overline{\Gamma_W}$ if the solution to this problem is larger than $\sqrt{1-\delta^2}$. In our case, the CVX package for specifying and solving convex programs \cite{cvx} was used to numerically solve the problem.

\bibliography{Refs}
\bibliographystyle{apsrev}

\end{document}